\documentclass[]{aa}
\usepackage{graphicx}

\begin{document}

\title{Planets in habitable zones:}

\subtitle{A study of the binary Gamma Cephei}

\titlerunning{Planets in Gamma Cephei}

\author{R.\ Dvorak, E.\ Pilat-Lohinger, B.\ Funk and F.\ Freistetter}
\authorrunning{Dvorak et al.}

\offprints{R.\ Dvorak, \email dvorak@astro.univie.ac.at}

\institute{Institute for Astronomy, University of Vienna, 
      T\"urkenschanzstrasse 17, A-1180 Vienna, Austria}

\date{Received; accepted}   

\abstract{

The recently discovered planetary system  in the binary $\gamma$  $Cep$  
was studied concerning its dynamical evolution. We confirm that the orbital parameters 
found by the observers are in a stable configuration. The primary aim of this study 
was to find stable planetary orbits in a habitable region in this system, which
consists of a double star (a=21.36 AU) and a relatively close 
(a=2.15 AU) massive ($1.7$ $m_{jup}$ $\sin i$) planet. We did straightforward numerical 
integrations of the equations of motion in different 
dynamical models and determined the stability regions for a fictitious massless planet  
in the interval of the semimajor axis $0.5$ AU $< a <$ $1.85$ AU around
the more massive primary. To confirm the results we used  the
Fast Lyapunov Indicators (FLI) in separate computations, which are a common tool for 
determining the chaoticity of an orbit.  Both results are in good agreement and 
unveiled a small island of stable motions close to 1 AU up to an
inclination of about $15^o $ (which corresponds to the 3:1 mean motion resonance 
between the two planets).
Additionally we computed the orbits of 
earthlike planets (up to 90 earthmasses) in the small stable island and found out, 
that there exists a small window of stable orbits on the inner edge of
the habitable zone in 
$\gamma$ $Cep$ even for massive planets.  
}

\maketitle

\section{Introduction}

Extra solar planets exist also in  double stars and, due to the fact that 
double and multiple star systems are more numerous
than single stars,  we expect many more discoveries in such systems
in future. All of the five  systems (table 1)
with  planetary companions are of the so-called S-type, where the planet orbits 
one primary.  This type of motion was studied theoretically since more than 
20 years (see e.g.\ Dvorak 1984 and 1986, Rabl \& Dvorak 1988, Holman \& 
Wiegert 1999, Pilat-Lohinger 2000a and 2000b, Pilat-Lohinger \& Dvorak
2002).
Stability studies have been undertaken for the 
other type of motion in binaries, the P-type, where the planet orbits both
components of a binary (see e.g. Goudas 1963, Dvorak et al.\ 1989, Holman \&
Wiegert 1999, Broucke 2001, Pilat-Lohinger et al.\ 2002), but up to now we do not have evidence that
they exist. From the cosmogonic point of view the S-types may
have formed similar to a planet around a single star.

\begin{table}[]
\begin{center}
\caption[]{Exoplanets in binaries}
\begin{tabular}{cccccc}
\hline\hline
Star & Planet & $M \sin i$ & a & e & period \\
&& [$M_J$]&[AU]&&[days]\\\hline
{\bf Gliese 86} & Gl86 b & 4 & 0.11 & 0.046 & 15.78\\
{\bf (K1 V)}&&&&&\\
\hline
{\bf 55 Cancri} & 55Cnc b & 0.85 &0.115 & 0.02 &14.653\\
{\bf (G8 V)}&55Cnc c & 0.21  &0.241 &0.339 & 44.275 \\
&55Cnc d &4.95  & 5.9& 0.16 & 5360 \\
\hline
{\bf 16 Cyg B} & 16CygB b &1.5& 1.72& 0.67 &804\\
{\bf (G2.5 V)}&&&&&\\
\hline
{\bf $\tau$ Bootis} & $\tau$Boo b & 4.09 & 0.05 & 0.& 3.312\\ 
{\bf (F7 V)}&&&&&\\
\hline
{\bf $\gamma$ Cephei} & $\gamma$Cep b & 1.76 &2.15 &
0.209 & 903\\
{\bf (K1 IV)}&&&&&\\
\hline
\end{tabular}
\end{center}
\end{table}

The recently discovered Jupiter size planet in HD 222404 ($\gamma$ $Cep$) (see Cochran et al, 2002)
has a slightly eccentric orbit in a mean distance of $a=2.15$ AU around
the K1 IV star (table 2). Our study shows the dynamical
stability of this planetary orbit with the published
orbital parameters. The main subject of 
our investigations 
is the existence of possible planets in the habitable region\footnote{According to
the common point of view the habitale zone is defined as the 
region around a star in which liquid water can exist on the surface of a 
planet (Kasting et al, 1993). Because of the spectral type of $\gamma$ $Cep$, the habitable
zone is in the region between 1 AU and 2.2 AU.} in a wider 
sense between $0.5$ AU $ < a < 1.85$ AU from the point of view of orbital dynamics.

\begin{table}[]
\begin{center}
\caption[]{The $\gamma$ $Cep$ planetary system }
\vspace{0.2cm}
\begin{tabular}{lccc}
\hline\hline
& Star A & Star B & Planet\\
\hline
Temperature [K] & 4900 & 3500 & ---\\

Radius [Solar Radii] & 4.7 & 0.5 & ---\\

Mass [Solar masses] & 1.6 & 0.4 & 0.00168\\

Period [years] &  & 70 & 2.47\\

Semi-major Axis [AU]&  & 21.36 & 2.15\\

Eccentricity &  & 0.44 & 0.209\\
\hline
\end{tabular}
\end{center}
\end{table}

\section{Numerical setup and dynamical model}

The equations of motion were integrated with two different
methods:
 
1. We used the Lie-integrator (see e.g. Hanslmeier and Dvorak 1984,
Lichtenegger 1984) specially adapted for these problems. It
uses an automatic step size and can quite precisely integrate also
high eccentric orbits. 
The time interval of computation  for the Lie-integration was set to 1 million years 
(which corresponds to approx.  $1.5 \times 10^4$ periods of the primaries). 
The stability definition for the orbits was the following:
An orbit is stable up to the moment when a ``possible crossing of the orbits'' occurs, which means
that the massive planet's periastron and the fictitious planet's apoastron 
allow close encounters. 

2. The Bulirsch-Stoer integrator
was used in connection with the computation of the Fast Lyapunov Indicators
(FLIs) (see Froeschl\'e et al.\ 1997); 
this is a well known tool to distinguish between chaotic and regular orbits.
The computations of the FLIs were carried out for $10^4$ binary periods. 
To determine the stable motion we defined a critical value for the FLIs
(i.e.\ $10^9$), up to which all orbits were found to be regular. 

The following dynamical models were investigated:

model {\bf A:} The simplest dynamical model, where one can  
explore the stability of 
a planet in a binary, is the elliptic restricted three body problem;
the planet's mass is neglected-- thus the binary describes 
an unperturbed Keplerian motion. This model was used for all studies cited
in the introduction above. 

model {\bf B:} In the ``restricted four-body problem''
(binary + massive planet  + massless fictitious planets)
we checked the orbits  in the region 
inside the orbit of the planet (0.5 AU to 1.85 AU)

model {\bf C:} The most realistic model for this system is a full 4-body problem 
where the inner -- fictitious -- planet  has also a mass, which we varied in the following
way: $m_{earth} \le m \le 90$ $m_{earth}$. 

In the framework of these three dynamical models we integrated the orbits of 
tenthousands of fictitious massless
planets (and hundreds of massive planets) for a different grid of 
semimajor axes and we varied
also the orbital inclination of the additional planet (the initial orbit was always
circular).

\section{The stability of the planet in $\gamma$ Cep}

Using the parameters of the binary given in table 2 we found for a massless 
regarded planet  a large
zone of stability (fig. 1). For different eccentricities of the binary and
also of the planet ($0< e < 0.5$) we computed 
 a diagramm of possible stable orbits for the mass ratio $\mu = m_B/(m_A+m_B) = 0.2$ .
The calculations for this study were done in the dynamical model {\bf A},
(the elliptic restricted three body problem) which does  not take into 
account a possible inclination of the planet's orbit. 
For the present configuration (the eccentricities of the binary $e = 0.4$ and the planet
 $e = 0.2$) 
the planet is in a very stable zone and could be stable up to almost twice its observed semimajor 
axes $a < 4$ AU. 
When we use theoretical results of the stability of P-types (see e.g.\
Pilat-Lohinger et al.\ 2002, and references therein)
we believe that even an inclined orbital plane of the planet's motion would 
not fundamentally change this diagramm.

\begin{figure}[h]
\centerline{
\includegraphics[height=8.5cm]{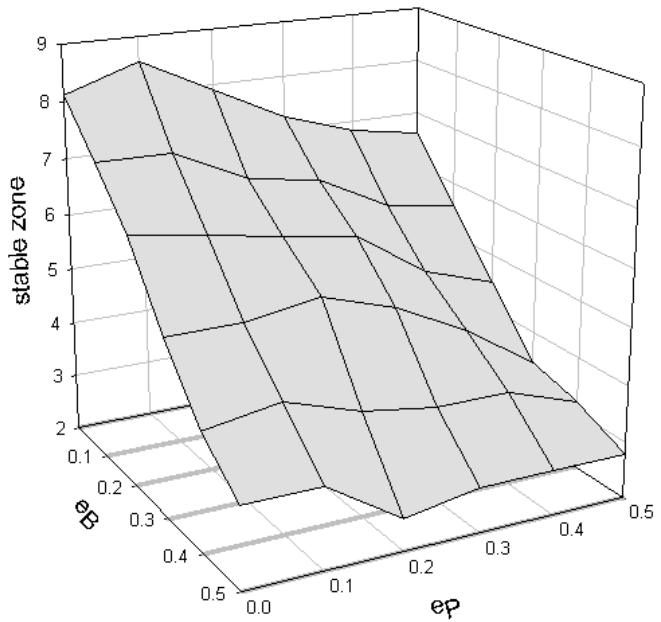}
}
\caption[]{The limits in AU of the stable zone for 
(z-axis) for the mass ratio $\mu = 0.2$ of $\gamma$ $Cep$, where the initial 
eccentricities of the planet (x-axis) and of the binary (y-axis) vary from 0 to 0.5}
\end{figure}

\section{Planets in habitable zones}

With a grid of $\Delta a =0.05$ between 0.5 AU and 1.85 AU we did computations of
the orbits of
fictitious massless planets in model {\bf B} and also took into account  possible
inclinations of the plane of motion of the planet in this region 
($\Delta i = 5^o$ for $0^o < i < 50^o$). Fig. 2 shows the escape regions 
according to their escape times determined with the Lie-integration. The dark regions 
are the ones were the planet escapes after a very short time and reaches
an eccentricity which allows close encounters between the ``real'' planet and the  
fictitious one, 
the white regions  show the stable orbits (within the time of integration). 
We can see the increase of the regions of stable orbits with larger inclinations, 
which is well understood from the geometry
of the different orbits which unables encounters for inclined orbits.
The most interesting feature is the stable region around 1 AU for the planar
orbits, which extends up to $i=10^o$, then disappears and reappears for larger inclinations
($i > 30^o$). The larger unstable zone between $40^o$ and $50^o$ is probably due to the acting 
Kozai resonance (Kozai, 1962).

\begin{figure}
\centerline{
\includegraphics[height=8.5cm]{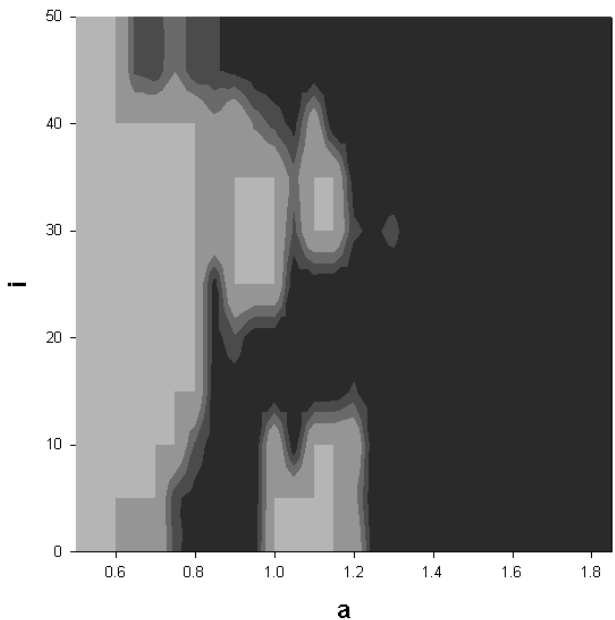}
}
\caption[]{Stability regions for $\gamma$ $Cep$; semimajor axis in AU (x-axis)
versus inclination. Only the light grey regions indicate stable orbits, the different
darker grey tones show the escape times in decreasing order; black
marks a fast escape within
10 periods of the binary}
\end{figure}

To check the r\^ole of the eccentricity of the known planet on the size of the stability regions, 
we made two more  computations in model {\bf B} in the planar problem: 
for $e=0.1$ we discovered 
an increase of the stable region up to $a=1.4$ AU. For $e=0.3$
the stable region extends only up to 0.65 AU without any stable window
in the unstable zone! 
Such an eccentric orbit does not allow a planet orbiting in a habitable region in $\gamma$ $Cep$. 

We repeated the study using the FLIs, but now we integrated for a slightly shorter time
because this chaos indicator is quite effective even for short time integrations. 
Because the method is also faster, we used a finer grid for the semimajor
axis ($\Delta a = 0.01$) and also for the inclination ($\Delta i = 1^o.7$). The
results of the FLIs shown in fig. 3 correspond quite well to fig. 2 
although more details are visible:
the stable region around 1 AU (3:1 resonance) is much smaller, but it also extends up
to an inclination of approximately $15^o$. This island of stability lies in a large
region of unstable motion, which becomes smaller with larger inclinations of the planets' orbit. The tiny islands of different grey indicate different values of the 
FLI.

\begin{figure}[h]
\centerline{
\includegraphics[height=8.5cm]{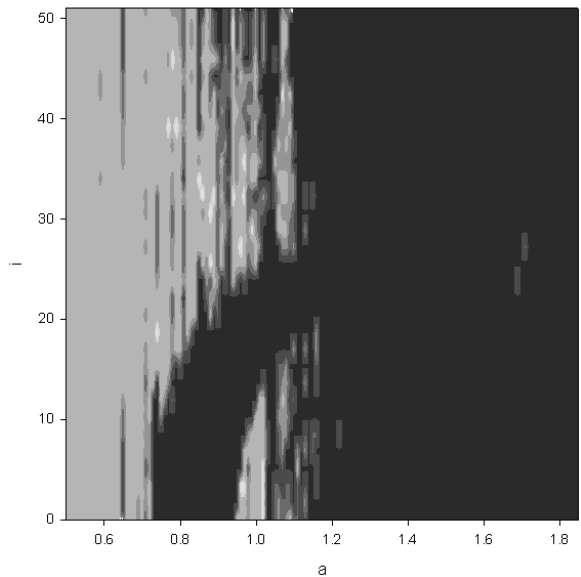}
}
\caption[]{Stability diagramm for $\gamma$ $Cep$ derived with the FLIs; semimajor axis in 
AU (x-axis) versus inclination. The different shades corresponds to different
values of the FLI, which splits up the phase space into different types of
motion: from light grey (stable) to black (chaotic). The latter corresponds to 
the unstable zone in fig.\ 2. }
\end{figure}

Additionally we checked with the Lie-integration
the region around 1 AU (for $i=0$) in the planar problem with separate computations 
for a finer grid ($\Delta a =0.02$). 
We confirmed the results of the FLIs (see fig. 3) which show the
complicated structure in this region close to the 3:1 mean motion resonance, 
where also high order resonances  and secondary resonances are present.

\section{Massive planets in the habitable zone}

The next problem to study was how a terrestrial planet may
influence the orbit of the discovered planet. We started the test
computation in the ``stable window'' which we discovered close to 1 AU
with masses of the fictitious planet $m_{earth} < m < 15$ $m_{earth}$ and with a grid of
$\Delta a =0.02$ AU but did not take into account a possible inclination of the fictive planets' orbit. 
In fig. 4 we show the results and indicate escaping orbits (dark),
unstable orbits (grey) and stable orbits (white). It is remarkable that only for
a = 1 AU all orbits for the different masses of the fictitious planet are stable. 
The comparison of the stability for a ''massless'' planet  with massive
planets shows, that there are differences, but the stability window is still present
although the structure changes. Test calculation for larger masses of the additional planet
close to  a = 1 AU showed, that there are still stable orbits up to m
= 90 earthmasses.   
Another interesting fact is, that
we found ``jumping orbits''; this means, 
that the semimajor axis jumps on a very small
scale irregularly between two different values (corresponding to very close high order
resonances inside the 3:1 mean motion resonance). This phenomenon was 
recently found for another extrasolar planetary system namely for HD 12661 
(see Kiseleva-Eggleton et al.\ 2002)
and is very probable due to a jumping between high order resonances as it was found by 
Milani et al. (1989).

\begin{figure}[h]
\centerline{
\includegraphics[height=7.5cm]{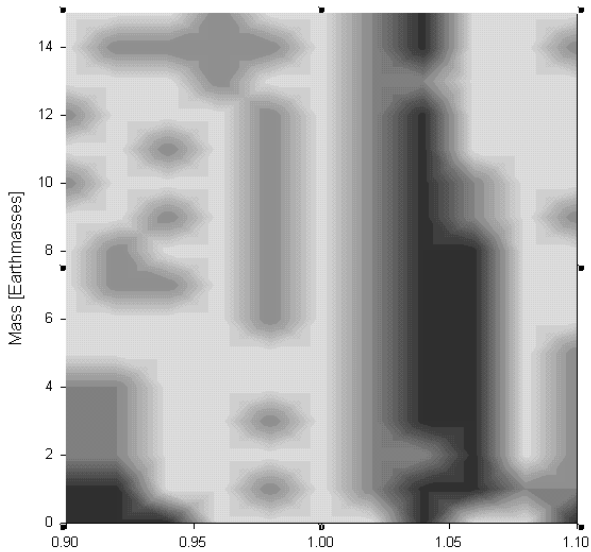}
}
\caption[]{Stability diagramm for $\gamma$ $Cep$ initial distance (x-axis) versus the
mass of a fictitious planet in Earthmasses (y-axis)}
\end{figure}

\begin{figure}[h]
\centerline{
\includegraphics[height=8.5cm,angle=270]{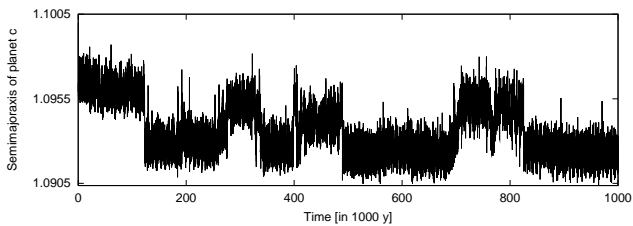}
}
\caption[]{A jumping orbit for a fictitious planet ($m=2m_{earth}$) in 
$\gamma$ $Cep$ for 1 million years}
\end{figure}

\section{Conclusions}

In this study we confirm the  dynamical stability
of the recently discovered planet in the binary $\gamma$ $Cep$ over at least some $10^8$ years, 
which let us assume that  the orbital parameters 
found by the observers are in a stable configuration up to cosmogonically time scales.
 Therefore the  primary aim of our investigations 
was to find stable planetary orbits in a habitable region in
this system. Due to the
results of extensive numerical 
experiments, for which we used direct integrations and also the FLIs, we discovered,  
that there could exist
planets in the stability regions depending on the inclination of the orbit 
of the fictitious planet  inside the orbit of the known planet. 
We unveiled the existence of a small island of stable motions 
close to 1 AU up to an
inclination of about $15^o $ (3:1 mean motion resonance 
between the two planets\footnote{This is opposite to our knowledge that the 3:1 resonance in the main 
belt of the asteroids is unstable. In a recent study by Hadjidemetriou
(2002) he also found that two 
planets in the 3:1 mean motion resonance around a single star do not have stable periodic orbits. 
Therefore we assume, that due to the action
of the second relatively massive primary, the motion of the fictitious planet is stabilized.}).
Final calculations with massive earthlike planets (up to 90 Earthmasses) in the small 
stable island confirm, 
that  even for massive planets there exists a small window of stable
orbits on the inner edge of the
habitable zone in $\gamma$ $Cep$.

\begin{acknowledgements}
E.\ Pilat-Lohinger wishes to acknowledge the support by the Austrian FWF
(Hertha Firnberg Project T122). B.\ Funk and F.\ Freistetter wish to 
acknowledge the support by the Austrian FWF (Project P14375-TPH).
Additionally we thank Drs. Kinoshita, Bois and Endl for their valuable advice.
\end{acknowledgements}

\end{document}